# Iterative Motion Compensation reconstruction ultra-short TE(iMoCo UTE) for high resolution free breathing pulmonary MRI

Xucheng Zhu[1,2], Marilynn Chan[3], Michael Lustig[1,4], Kevin Johnson[5,6], Peder Larson[1,2]

*1. UCSF/UC Berkeley Graduate Program in Bioengineering, University of California, San Francisco, San Francisco, California, United States*

*2. Department of Radiology and Biomedical Imaging, University of California, San Francisco, San Francisco, California, United States*

*3. Department of Pediatrics, Division of Pediatric Pulmonology, University of California, San Francisco, San Francisco, California, USA*

*4. Department of Electrical Engineering and Computer Sciences, University of California, Berkeley, Berkeley, California, United States*

*5. Department of Medical Physics, University of Wisconsin, Madison, Madison, Wisconsin, United States*

*6. Department of Radiology, University of Wisconsin, Madison, Madison, Wisconsin, United States*

Corresponding author:
Peder E.Z. Larson
1700 4th St, Room 102C
San Francisco, CA 94143
Tel: (415) 415-4876
Fax: (415) 415-4451
E-MAIL: peder.larson@ucsf.edu

Word count: 4358

**Abstract:**

**Purpose:** To develop a high scanning efficiency, motion corrected imaging strategy for free-breathing pulmonary MRI by combining a motion compensation reconstruction with a UTE acquisition, called iMoCo UTE.

**Methods:** An optimized golden angle ordering radial UTE sequence was used to continuously acquire data for 5 minutes. All readouts were grouped to different respiratory motion states based on self-navigator signals, then motion resolved data was reconstructed by XD Golden-angle RAdial Sparse Parallel reconstruction (XD-GRASP). One state from the motion resolved images was selected as a reference, and motion fields from the other states to the reference were derived via non-rigid registration. Finally, all motion resolved data and motion fields were reconstructed by using an iterative motion compensation reconstruction with a total generalized variation sparse constraint.

**Results:** The iMoCo UTE strategy was evaluated in volunteers and non-sedated pediatric patient(4-6 y/o) studies. Images reconstructed with iMoCo UTE provided sharper anatomical lung structures, and higher apparent SNR and CNR, compared to using other motion correction strategies, such as soft-gating, motion resolved reconstruction, and non-rigid motion compensation(MoCo). iMoCo UTE also showed promising results in an infant study.

**Conclusions:** The proposed iMoCo UTE combines self-navigation, motion modeling, and a compressed sensing reconstruction to increase scan efficiency, SNR, and reduce respiratory motion in lung MRI. This proposed strategy shows improvements in free breathing lung MRI scans, especially in very challenging application situations, such as pediatric MRI studies.

**Introduction**

MRI has the potential to assess pulmonary diseases by providing soft-tissue contrast and structural information within the lung[1]. Compared to CT scanning, MRI avoids ionizing radiation exposure, which would be safer for pediatric subjects[2], or patients requiring longitudinal follow-up imaging[3]. However, pulmonary MRI is challenging due to short T2*, low proton density of the lung parenchyma[4,5] and subject motion, especially respiratory motion[6].

Ultra-short echo time(UTE)[7] and zero echo time(ZTE)[8] type acquisition strategies have been developed which preserve short T2* signal in the lung by providing the means to collect images with a sub-millisecond echo time(TE). Such sequences use optimized excitation pulses and readouts strategies to maximize SNR for pulmonary imaging[9], however, most UTE/ZTE sequences still take a few minutes or longer, usually 5 to 10 minutes[5,10,11], required to obtain sufficient lung parenchyma SNR with high spatial resolution. During the scan, there are inevitable motion effects, especially from respiratory motion. In addition, longer scan time increases the possibility of irregular motion of the subject, especially for pediatric subjects[12] and subjects with poor pulmonary function.

A variety of respiratory motion compensation strategies have been developed, most of which utilize motion tracking for retrospective motion correction or compensation. External respiratory belts are widely used to indirectly track the motion by measuring the respiratory-induced abdomen stretching. An alternative way is to use the repeatedly acquired k-space center (DC), which is feasible for center-out UTE sequences, measuring the signal change caused by respiratory motion[7,11,13,14]. To more accurately and directly characterize subject motion, low spatial but high temporal resolution 2D/3D images could be reconstructed and used as a self-navigator [6,15–17].

Most of the motion correction and compensation methods can be classified into three categories. 1) Respiratory gating: based on the respiratory motion signal, only data acquired within a certain motion state, usually the end expiratory state, is used for reconstruction[18]. However, gating based methods reduce scan efficiency and prolong the scan time. To increase the scan efficiency, the soft-gating method was proposed to add non-zero weightings on the data[6,19,20], but it reduces the ability to correct motion[10].

2) Motion resolved reconstruction: instead of reconstructing a single motion gated image, all acquired data are grouped in different motion states, and spatial correlation of the different images are used as the prior information for compressed sensing based reconstruction, such as XD Golden-angle RAdial Sparse Parallel reconstruction (XD-GRASP)[21,22], kt- FOCal Underdetermined System Solve (kt-FOCUSS)[23]. 3) Motion compensation (MoCo) reconstruction with image registration: Unlike gating or soft-gating strategies, motion compensation strategies align all motion states images to the same state via image based registration. After registration, all motion states images are summed to compose a single image, therefore increasing the data acquisition efficiency[24]. MoCo type strategies have been applied in simultaneous PET/MR applications for increase PET image SNR[25–29]. A more sophisticated way that Batchelor[30] first proposed is the generalized matrix description(GMD), to formulate the motion deformation as a matrix operator to describe the motion propagation in signal space. A few motion compensation reconstruction strategies[31–35] based on GMD have been proposed and applied to cardiac MRI.

In this work, we propose a new free breathing motion corrected pulmonary MRI strategy, iterative Motion Compensation reconstruction ultra-short TE, called iMoCo UTE, to improve high spatial resolution free breathing lung MRI. There are three main components of the proposed method, 1) a pseudo random non-Cartesian UTE sequence, 2) motion resolved reconstruction and motion estimation, and 3) a novel iterative motion compensation(iMoCo) reconstruction with compressed sensing. The iMoCo reconstruction iteratively fits the data to a non-rigid motion model, and leverages compressed sensing principles to further suppress noise and artifacts. When combined with a UTE sequence, the proposed method addresses the challenges in lung MRI of intrinsically low SNR and motion by providing a high acquisition efficiency and motion robust pulmonary MR images, especially in challenging situations, such as pediatric MRI studies. The proposed strategy is evaluated on both healthy volunteers and pediatric patients, and compared to other motion correction strategies, such as soft-gating, motion resolved reconstruction, and image based motion compensation(MoCo) strategies.

**Methods:**

**Overview workflow of iMoCo UTE**

The overall workflow of proposed iMoCo UTE is summarized in Figure 1. Data are acquired with a pseudo randomly sampled 3D radial UTE sequence which results in k-space sample ordering being uncorrelated with the respiratory motion. Then, data are binned to different motion states according to respiratory tracking signals. Respiratory tracking is first determined using a k0/DC navigator signal. Then, additional image based navigator is used to remove the data with irregular motion(Figure 1.a). Instead of doing high spatial resolution motion resolved reconstruction, the motion resolved data is used first to reconstruct ~1.5 times coarser spatial resolution motion resolved images(Figure 1.b). Motion fields from different motion states to a reference state then are estimated via non-rigid registration[36](Figure 1.c). Finally, the estimated motion fields and a spatial total generalized variation(TGV) sparsity constraint[37] are added into the iterative motion compensated reconstruction(iMoCo) model to reconstruct a single state high resolution motion-free image.

**UTE sequence**

An optimized 3D UTE sequence with slab selection and variable density readout acquisition[9], which increases the SNR efficiency and reduces aliasing artifacts, was used for pulmonary imaging scans. A golden angle ordering acquisition scheme was used to randomize undersampling artifacts over time and improve the motion resolved reconstruction and motion estimation. All the studies were performed on 3T MRI clinical scanners (MR750, GE Healthcare, Waukesha, WI, USA). More specific acquisition parameters are listed in the experiment section.

**Respiratory motion detection and motion resolved reconstruction**

A DC based self-navigator was used for initial respiratory motion detection, in which the first points of the radial readouts are used as self-navigator signals. The multi-channel data were combined via the adaptive navigator strategy[38]. Then DC signals were filtered by low-pass (0.5~1 Hz cut-off frequency) filter to reduce high frequency noise. The DC navigator was used as long as no large fluctuation of the DC signal (defined as having a signal deviation from base line that was 3 times greater than range of the respiratory signal) was observed. If there was a baseline fluctuation of the DC signal, alternatively, a 3D image based navigator, generated by a locally low rank constrained reconstruction[6], was used to identify and discard bulk motion corrupted data. The respiratory motion signal was derived from the DC signal after discarding the irregular motion corrupted data and baseline correction. (The 3D image navigator was not used for respiratory motion estimation due to limitations on the spatial (~3mm isotropic) and temporal resolution (~300ms) that were not sufficient to capture pediatric respiratory motion.)

Based on the motion estimates, the acquired data was binned into different respiratory motion states for motion resolved reconstruction, a process similar to XD-GRASP[21]. Even when data are binned into different motion states, there is inevitable residual motion among data within the same motion state. To minimize this effect, a large number of motion states, between 8 and 10, was used. This is in contrast to standard XD-GRASP, which typically uses 4 to 6 motion states[22]. However, since the full dataset(all spokes) are usually 3 to 4-fold undersampled, if number of motion states increases to 8 or higher, the undersampling factor would go up to 30, which would lead to strong streaking artifacts and lower SNR even with compressed sensing and parallel imaging. To accommodate the higher undersampling and to reduce the reconstruction artifacts, a coarser resolution(~1.5 times the native resolution) was used, which maintains a reasonable undersampling factor in this workflow. Since the purpose of the binning is for estimating motion between the bins and correct for it, a full resolution reconstruction is not necessary. Reducing the resolution has the added benefit of reducing the computational load. The reconstruction was done via an XD-GRASP type reconstruction solving,

$$\underset{X}{\mathrm{argmin}} \sum_{i,k}^{N,m} \left\| W(FS_i X_k - d_{ik}) \right\|_2^2 + \lambda_s \left\| \Phi X \right\|_1 + \lambda_t TV_t(X) \ . \tag{1}$$

Here, the squared-error data consistency term (left) includes multi-channel sensitivity maps $S_i (i = 1,2, \dots N)$, motion states sorted multi-channel data $d_{ik}$, $W$ is sampling density compensation weights, $F$ is the non-uniform Fourier transform operator, implemented via gridding algorithm, and $X_k$ are the motion-state 3D images, parametrized by motion state index $k$. A spatial sparsity 3D l1-wavelet term, where $\Phi$ is the wavelet transform, and a motion dimension total variation term, $TV_t()$, are added to the reconstruction. Additional motion-resolved reconstruction formulations including removing the spatial l1-wavelet term or replacing it with spatial total variation were also evaluated, as described in the Supporting Information Figure S1. Following the motion resolved reconstruction, one of the motion states (e.g. typically the end expiratory state) was selected as a reference frame. All other motion state images were then registered to the reference via Demons non-rigid registration[36] (4 pyramid levels coarse-to-fine registration and 100 iterations are used in Demons). Estimated motion fields were interpolated to match the full resolution image, and used in the following motion compensation reconstruction.

**Iterative motion compensated reconstruction (iMoCo)**

Once the motion fields were derived, the entire data is used to reconstruct a single frame, that is motion corrected. We leverage the relation that $M_k \hat{X} = X_k$, where $M_k (k = 1,2 \dots m)$ are derived motion fields, $\hat{X}$ is the final reconstructed image, to include our motion estimates in the forward model. As described in the GMD model, the motion field can be formulated as a linear operator. Although the inverse operator of non-rigid deformation is difficult to calculate, the adjoint operator can be simply estimated as reverse deformation from the reference to a certain state image. To further reduce streaking artifacts caused by undersampling and residual motion, a spatial total generalized variation (TGV) sparsity regularization term is added to the model. Unlike TV regularization, TGV relaxes the assumption that image is piecewise constant, which would be more suitable for continuous signal changes of tissues[37]. So the reconstruction problem can be reformulated as optimization problem:

$$\underset{\hat{X}}{\mathrm{argmin}} \sum_{i,k}^{N,m} \left\| W(FS_i M_k \hat{X} - d_{ik}) \right\|_2^2 + \lambda_s \mathrm{TGV}_s(\hat{X}) \ . \tag{2}$$

In the data consistency term(left), $I$, $d_{ik}$, and $F$ are the same notations as the motion resolved reconstruction(Eq. 1) with the addition of derived motion fields $M_k (k = 1,2 \ldots m)$. $\hat{X}$ is the final reconstructed, single state high resolution image. The sparse penalty term(right) is $\text{TGV}_s()$. The optimization problem is solved by using first order primal-dual algorithm[39].

**Experiments**

All human studies conducted were approved by UCSF Institutional Review Board (IRB). Several different types of studies are included in the results and discussions.

For adult healthy volunteer studies(N=7), scan parameters included: prescribed field of view(FOV) = $32 \times 32 \times 32 \text{cm}^3$ ($64 \times 64 \times 64 \text{cm}^3$, 2-fold oversampling on readout direction for radial sequence), flip angle = $4°$, 1.25mm or 1mm isotropic resolution, readout bandwidth = $\pm 125 kHz$, $TE = 70\mu s$, $TR = 2.7 - 3.1 ms$, $TR$ increased as the prescribed FOV was reduced. The total scan time was approximately 5 min to 5 min 30 s. Number of total acquired spokes of each scan was approximately 100,000.

For pediatric patients diagnosed with pulmonary diseases studies(N=4), the FOV was prescribed based on the size of the subject (22~26cm), and spatial resolution (isotropic) was kept to 1.1mm or higher(<1mm) due to their smaller anatomical structure size. Both $TE$ and $TR$ ($TE = 80\sim110\mu s$, $TR = 3.1\sim3.7ms$) increased compared to adult studies due to smaller FOV and excitation slab. The number of spokes was adjusted between 80,000 to 90,000 to keep the total scan time no more than 5 min 30 s. Due to varied scan subject size, different receiver coil arrays were used in pediatric scans to improve SNR: 8-channel and 32-channel cardiac arrays (GE Healthcare, Waukesha, WI, USA), and a 12-channel flexible screen-printed coil array (Inkspace Inc., Moraga, CA, USA)[40].

For the infant patient study(N=1), the following scan parameters were used. FOV was 18cm, spatial resolution (isotropic) was 0.9 mm, $TE = 170\mu s$, $TR = 4.6ms$), the number of spokes was 75,000, and total scan time was around 5 min 30 s. 8-channel head coil (GE Healthcare, Waukesha, WI, USA) was used in the experiment due to very small subject size.

**Data processing and imaging reconstruction**

The self-navigator, motion signal processing, and iMoCo reconstruction were implemented in MATLAB (Mathworks, Natick, MA). Coil sensitivity maps calibration, motion resolved reconstruction and soft-gating reconstructions were carried out by the Berkeley Advanced Reconstruction Toolbox (BART)[41]. All the quantitative measurements were also implemented in MATLAB.

**Image quality comparison and evaluation**

For image quality comparisons, we implemented non-gating, soft-gating, motion resolved, and non-rigid motion compensation (MoCo) reconstruction and motion correction strategies following the details in previous works[6,19,22]. Briefly, the MoCo method used the motion resolved reconstruction from XD-GRASP to genereate images across motion states, which were then non-rigidly registered using Demons and then averaged together. Hyperparameters used were kept the same in all the reconstructions.

To quantitatively compare the image quality among different motion correction strategies, several image metrics were computed. The sharpness of the lung-liver interface or diaphragm was measured via the relative maximum derivative(MD), defined as the maximum intensity change between lung-liver interfaces divided by mean intensity in the liver.

Apparent signal-to-noise ratio (aSNR), defined as the average signal over a small region of interest (ROI) divided by standard deviation of area out of the subject, was measured. Three representative areas were selected as ROIs: an airway, lung parenchyma, and the aortic arch. Instead of SNR, aSNR was measured in this work because the reconstruction methods could introduce spatially varying noise and also perform inherent denoising. Images reconstructed with different strategies were spatially aligned, so ROIs at the same locations could be manually drawn on all the reconstructed images for measurement.

Contrast-to-noise ratio (CNR), defined as the contrast difference over noise level, was also computed. In lung MRI, it is valuable to distinguish air, lung parenchyma, and vessels. Therefore, CNR between lung parenchyma and air, and between aortic arch and air were measured.

A paired sample t-test (p<0.05)  method was used for statistical comparison of the quantitative measurements MD, aSNR, and CNR.

## Results:

### Volunteer studies comparison

Volunteer study results are shown in Figure 2. One sagittal slice image from two adult volunteer studies with the proposed iMoCo and other motion correction strategies are shown in the first rows. The overall image quality of both subjects with iMoCo method is better than motion resolved and soft-gating methods, due to higher data usage efficiency and a more accurate motion model. The zoomed-in images show that the proposed reconstruction has better lung parenchyma contrast and less residual motion compared to the soft-gating method, and sharper edges of airways and vessels compared to the MoCo method.

To compare the capability to visualize small vessels and airways in the lung, two higher spatial resolution(1mm isotropic resolution versus 1.25 mm isotropic resolution in Figure 2) examples are shown in Figure 3. One coronal slice from each study is shown in the first row, and a maximum intensity projection(MIP) of 30 slices centered at the first row coronal images positions are plotted in the second row. Images with the iMoCo reconstruction have the best visual image quality, and show lower noise level, sharper vascular structures and more small blood vessels compared to other methods.

### Pediatric patient studies

Pediatric pulmonary MRI studies are much more challenging, especially for non-sedated free breathing scans. First of all, it is difficult for children to keep still during a long scan. Also, their respiration rates tend to be higher and less regular. In addition, the quiescent period after exhalation is much shorter, which might reduce the SNR efficiency and image quality of gating and soft-gating based methods[10]. Pediatric patients with different types of lung diseases and of different ages were scanned to show the capability of imaging different lung abnormalities with iMoCo UTE. Results of 3 representative pediatric studies with different observed abnormal lung structures are shown in Figure 4.  An image slice with the abnormality is shown in the first row, and a zoomed-in image in the second row. Images carried out by different reconstruction algorithms are compared for 3 patients: Patient 1 was a 5 year old female who had a severe combined immunodeficiency (SCID) post stem cell transplant with several observed lung nodules (red dashed circles); patient 2 was a 4 year old male with systemic juvenile idiopathic arthritis and childhood interstitial lung disease (chILD) with observed ground-glass opacity (green dashed circle, this opacity was also observed on CT); and patient 3 was a 8 year old female with surfactant protein C deficiency who had small pneumatoceles (lung cysts, blue arrow).  Images with the iMoCo reconstruction had the best depiction of these pathologic features. Particularly in patient 3, the pneumatocele has a much sharper boundary and better contrast with iMoCo compared to other methods.

**Hyperparameters selection**

In the proposed iMoCo reconstruction Eq. 2, there are two tunable hyperparameters, one is number of motion states, and the other is the TGV regularization weighting. Experiments with different hyperparameters were designed to investigate the selection of hyperparameters for the reconstruction.

iMoCo reconstructions carried out using different numbers of motion states are compared in Figure 5. In the example, the diaphragm motion range was ~1cm, the motion resolved spatial resolution was 1.5mm, and the final resolution was 1.25mm.  As the number of motion states increased from 2 to 6, the reconstructions improved, especially close to diaphragm(green arrow), however, trivial improvement was observed as the number increased from 6 to 8. We also quantitatively evaluated the effects from number of motion states on the final motion correction performance. The diaphragm maximum derivative(MD) were used to represent the motion correction performance, plotted in (b). Higher MD means better motion correction result. As number of states increase 2 to 6, 4 out of 4 cases show MD increase. As number of states goes larger, the improvement is inconsistent among different cases. Therefore, we believe the number of states could be estimated by the motion range divided by the motion resolved reconstruction resolution (in this example, 1cm / 1.5mm = 6.6). 8 motion states were used in all the results in this work.

In Eq. 2, $\lambda_s$ is the hyperparameter tuned to control spatial TGV regularization term. The reconstruction with TGV regularization will reduce the noise, and suppress undersampling and residual motion artifacts. Reconstructed images with different $\lambda_s$ are shown in Figure 6, using image data from one of the high resolution studies shown in Figure 3. As $\lambda_s$ increases, the noise and streaking artifacts are reduced. However, as $\lambda_s$ increases to 0.1, the overall images look over-smoothed, especially as some small vessels are blurred out, seen in the green circle area in Figure 6(a). MIP of 30 slices are shown in (b) to further compare the effect of $\lambda_s$ on small structures. $\lambda_s = 0.05$ shows less noisy without sacrificing the small vessels structures. aSNR measurements are used to present quantitatively present the effect of $\lambda_s$, where Supporting Information Figure S3 shows the aSNR change from airway, lung parenchyma and aortic arch as $\lambda_s$ increases. Although the all the aSNRs increase along $\lambda_s$, aSNR in airway is expected to be 0, which indicates that $\lambda_s$ goes to 0.1 or higher is over-regularized. Therefore, $\lambda_s$ was set to 0.05 for the iMoCo reconstructions.

**Quantitative measurement**

For all volunteer(n=7), and pediatric patient(n=4) studies, we quantitatively measured the MD on 10 sagittal slices with different motion correction strategies, and normalized the MD to the mean liver signal intensity close to the diaphragm. Measurements are summarized in Figure 7. Images carried out by iMoCo methods show significantly higher MD value compared to non-gating, soft-gating ,and MoCo methods. 6 out of 11 subjects had the highest diaphragm MD with iMoCo reconstruction.

We also measured the aSNR and CNR of certain regions in the lung, and the results are summarized in Figure 8. The aSNR in airways is expected to be close to 0 since there is very little $^1H$ density in air, and aSNR in lung parenchyma and aorta approximate the SNR level of short and long T2* tissues, respectively, with the reconstruction methods. The aSNR increase in MoCo images can be attributed to the lower apparent spatial resolution due to smoothing effects induced by the deformation interpolation, which can be observed in the image results in Figure 2-4. Images with iMoCo reconstruction have higher lung parenchyma and aortic arch aSNR, compared to motion resolved and soft-gating reconstructions. In addition, the airway aSNR with iMoCo is relatively low, which would benefit distinguishing airways from lung parenchyma. In Figure 8(c), iMoCo has significant higher CNR of the lung parenchyma and aortic arch, compared to soft-gating and motion resolved reconstructions. These quantitative measurements indicate that iMoCo method can not only achieve higher aSNR, but also reduce respiratory motion artifacts, which would benefit the relatively low SNR found in lung MRI.

**Feasibility of infant study**

The significance and feasibility of infant and neonatal lung MRI studies has been reported in previous work[12,42,43]. High spatial resolution and sufficient SNR are required to visualize smaller structures of the lungs. Figure 9 shows one 5-minute UTE scan of an unsedated 10-week-old infant with Pulmonary Interstitial Glycogenosis(PIG) with 0.9mm isotropic resolution case. Vessel structures(red arrow) and airways(green arrows) are largely improved with iMoCo reconstruction compared to other methods. An image based navigator[6] was also used in this study to capture the bulk motion of the baby, then the data corrupted by bulk motion were rejected, shown in (b). Image results with bulk motion rejection shows sharper vessels and diaphragm(red arrows) compared to without bulk motion rejection, in (c). iMoCo with the same hyperparameters were used for both of the Figure 9(c) reconstructions.

**Discussion:**

In this work, we proposed a new motion correction strategy by combining UTE, motion compensation, and a compressed sensing reconstruction to achieve high resolution free breathing lung MRI, called iMoCo UTE. Although soft-gating type reconstruction strategies have been widely used, they inevitably suffer from the residual motion in the reconstruction data. As the desired reconstructed spatial resolution goes higher, fine structure cannot be reconstructed due to this residual motion. The other option is to use a motion resolved type reconstruction, where all data are binned to different motion states then using spatial similarity across different motion states to reduce the downsampling artifacts. As the number of bins increases, on one hand, the residual motion would reduce, however, on the other hand, a higher undersampling factor in each bin would induce undersampling artifacts. iMoCo, aims to model the respiratory motion effects, and incorporate spatial motion compensation instead of directly weighting data or segmenting data. Through volunteers and pediatric patient studies, iMoCo shows capability to achieve high resolution, high SNR lung images without inducing motion artifacts. According to the quantitative comparison, the iMoCo and motion resolved reconstructions have the highest MD, and iMoCo and MoCo have higher aSNR compared to other methods. In addition, iMoCo is a general motion correction and reconstruction framework, so it could be extended to other applications, and is compatible with different trajectory designs.

**Motion resolved reconstruction and motion fields estimation**

As mentioned in the results, the number of motion states in motion resolved reconstruction determines the intra-bin residual motion, which would significantly affect the final reconstruction. However, iMoCo also largely depends on the motion field estimation. More motion states might degrade the motion resolved reconstruction images, leading to image registration errors. iMoCo also depends on accurate motion field estimation, and errors in the motion field would propagate to final reconstructed image causing blurring or ghosting artifacts.

Previous UTE lung studies have reconstructed 4-6 motion states in motion resolved reconstruction with an approximately 2-fold undersampling rate overall that resulted in around 10-fold undersampling rate in each motion state data[6,22]. For the motion resolved reconstruction this work, we increased the number of motion states to reduce the intra-bin motion and also used a slightly coarser spatial resolution to reduce the undersampling rate. This was aimed to ensure that the reconstructed image quality would not degrade the motion fields estimation and following motion compensated reconstruction. We set the motion resolved reconstruction to 1.5mm isotropic resolution to keep the undersampling rate in single motion state under or around 10, as the number of motion states was set up to 12, while providing images that could distinguish inter-bin motion as small as 1.5mm. In addition, we also compared different image registration methods, and Demons showed the best performance among the methods. An example is showed in Supporting Information Figure S2.

The regularization terms and values for the motion-resolved reconstruction in Eq. 1 were chosen empirically based on qualitative evaluation. We also evaluated the effects on motion field estimation by using different regularizations in motion resolved reconstruction, summarized in the Supporting Information Table S1, and Table S2. We calculated the RMSE of two different motion state images before and after registration, and the linear correlation coefficient and mean Euclidean distance between motion fields derived from different motion resolved reconstructions. The results were very similar between the different choices of regularization, indicates that the motion estimation is insensitive to the choice of regularization terms. There also maybe room to improve the proposed method through more sophisticated techniques, such as locally low rank contraints[44], which also could improve the overall iMoCo method.

In this work, we used one-to-one image registration between pairs of motion state images, and registration error might be further reduced by using group-wise registration. All the motion states images could be registered simultaneously, which might improve the robustness of registration process by taking advantage of the registration correlation among the motion states, and some work has shown promising results in 4D CT applications[45].

**3D navigator and irregular motion handling**

In this work, the self-navigator signals are simply used for respiratory motion binning. By using more sophisticated techniques, a 3D image based navigator could be extracted, then used to detect the bulk motion or irregular motion. In this work, data contaminated with bulk motion are discarded[6,12], which reduces the scan efficiency. We might also be able to incorporate bulk motion or even more complicated irregular motion into iMoCo reconstruction, which might increase the scan efficiency and robustness to irregular motion.

**Respiration related pulmonary abnormalities**

Some pulmonary abnormalities, such as air trapping, might lead to intensity changes during respiration[46]. One of the limitation of our studies is that iMoCo did not include signal intensity changes in the models, which may lead to missing the dynamic change of the abnormalities during respiration.

## Conclusions

In this work, we proposed a new free breathing high resolution pulmonary MRI strategy, combining motion compensation, UTE, and compressed sensing, called iMoCo UTE. iMoCo UTE has been validated and evaluated via both volunteers and patient studies, and shows potential in pediatric and infant pulmonary MRI studies.

## Acknowledgement

The authors would like to acknowledge Frank Ong, Ph.D., Scott K Nagle, M.D./Ph.D., for helpful discussion. The work is supported by NIH grant R01 HL136965.

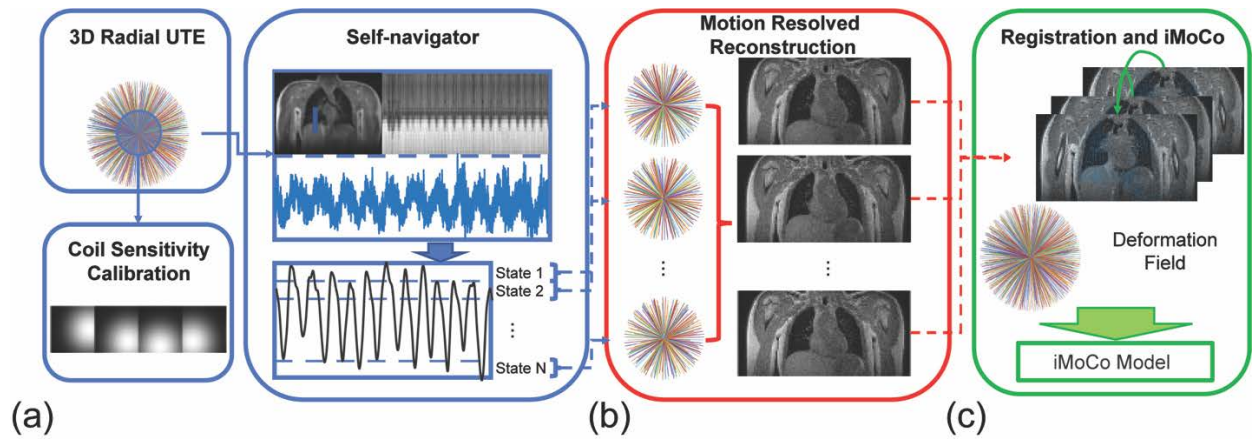

Figure 1. Overview of the iMoCo UTE workflow. (a) After optimized 3D radial UTE data are acquired, the center k-space is cropped out for coil sensitivity calibration, then used to estimate the respiratory motion signal via the k0 signal or a reconstructed 3D image navigator. (b) Based on the self-navigator signal, data are grouped in different motion states, then motion resolved images with medium spatial resolution are reconstructed. (c) All motion states images are registered to the selected reference state image via non-rigid image registration, then the deformation fields and the whole dataset are fed in the iterative motion compensation reconstruction model.

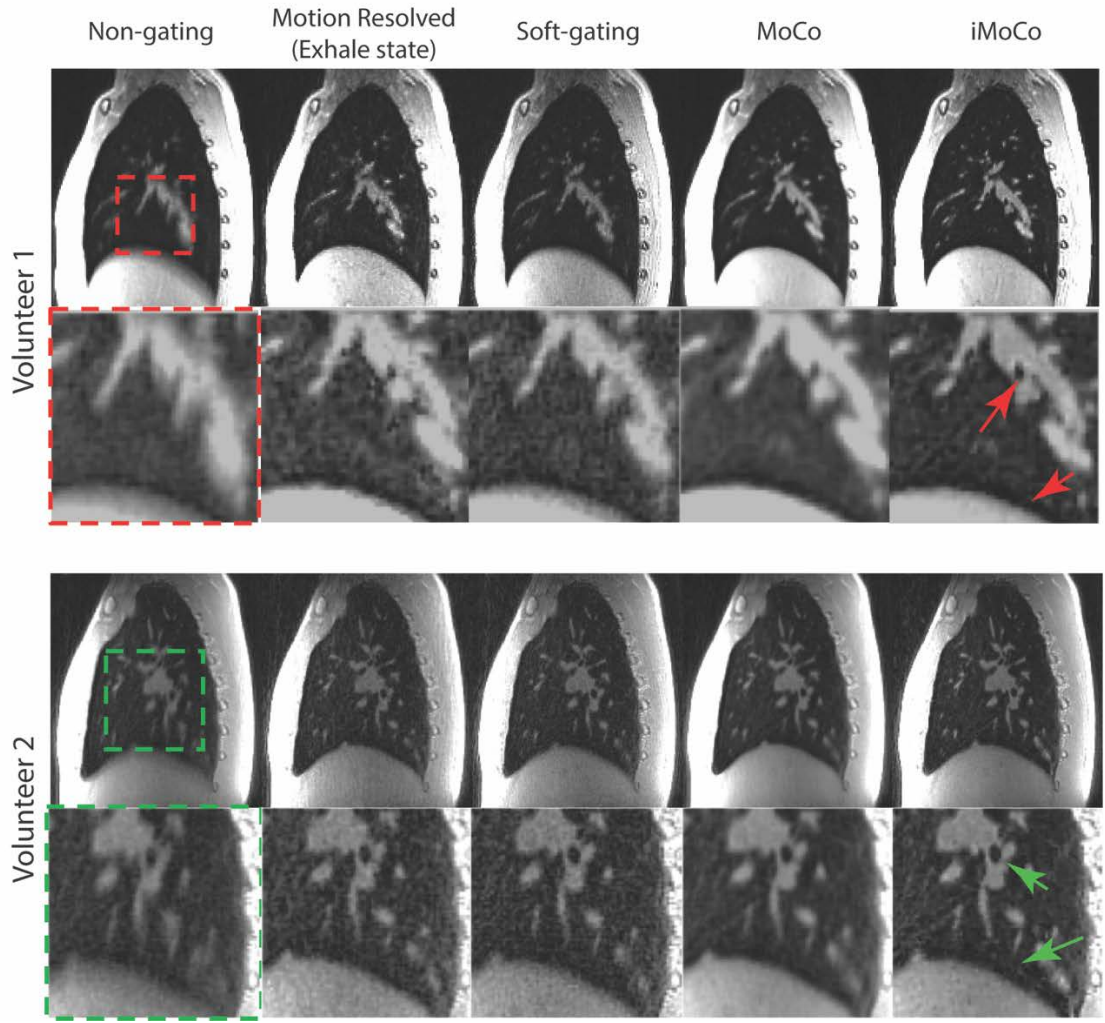

Figure 2. Example of volunteer study image results. The same slice images of each subject reconstructed with different motion correction and reconstruction strategies are plotted in the first rows, and a dashed square targeting at the center area of the lung are zoomed in and plotted in the second rows. In both cases, the iMoCo strategy had the sharpest image features and highest apparent SNR.

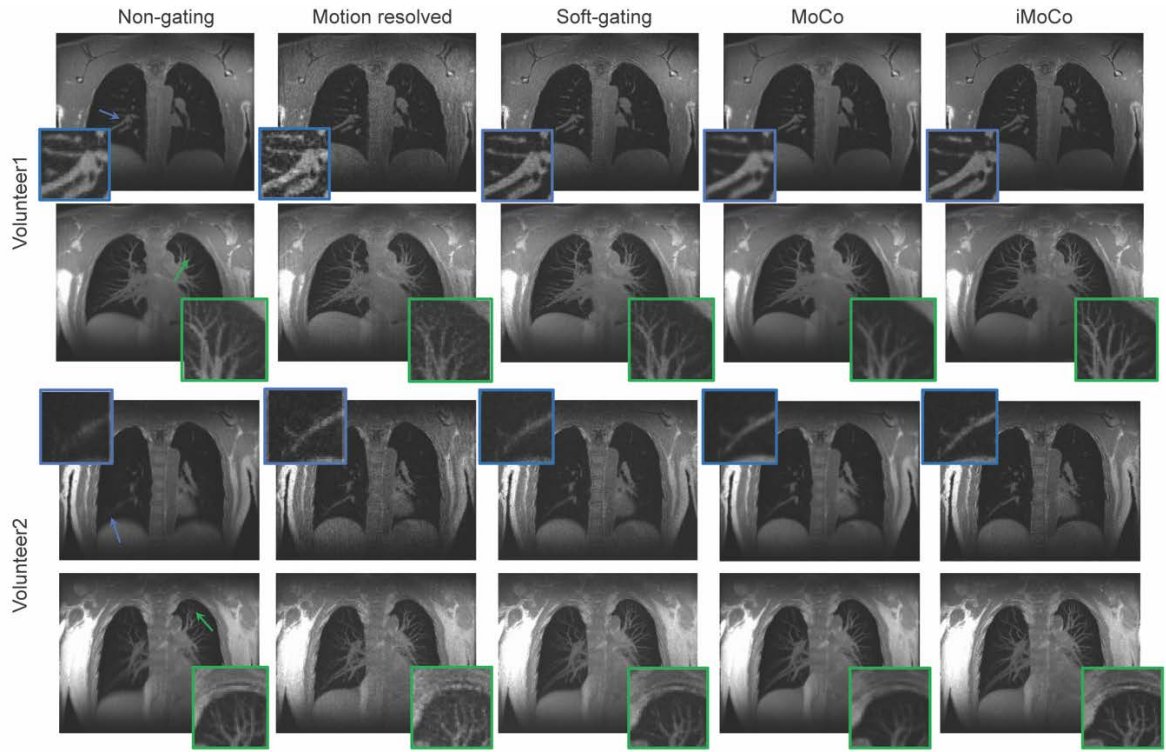

Figure 3. Example of high spatial resolution (1mm isotropic) lung images. The proposed iMoCo reconstruction is compared to non-gating, motion resolved reconstruction(exhale state), soft-gating and motion compensation(MoCo) reconstructions. In each volunteer example, the first row shows the one slice from different reconstructed images, and the second row shows the maximum intensity projection(MIP) of 30 slices in the AP direction. The area pointed with the arrows in the non-gating images are zoomed in. The iMoCo reconstruction was able to delineate the smallest pulmonary vessels.

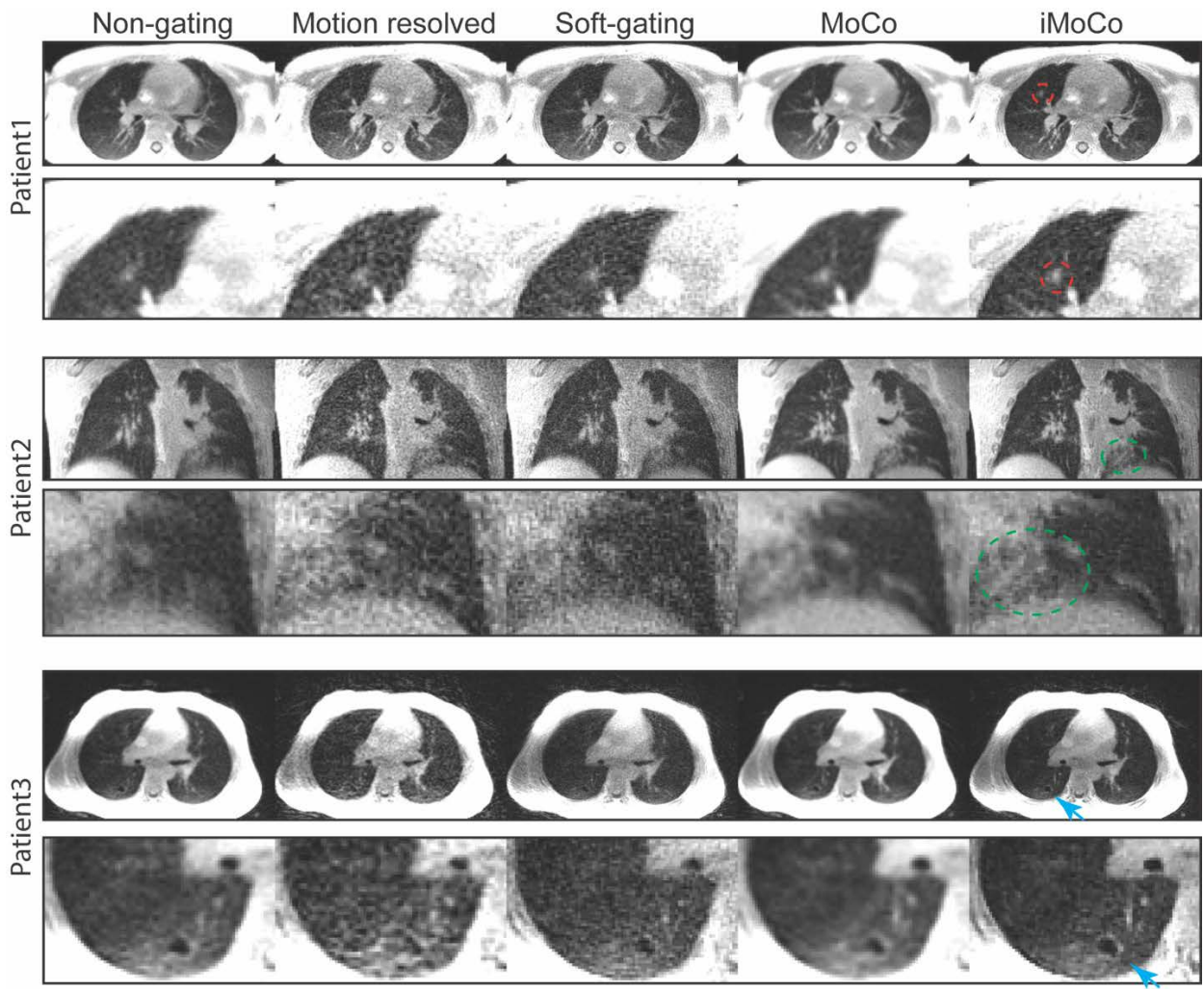

Figure 4. Pediatric patient study examples. Three different patient lung UTE scan(1mm isotropic resolution) results with different motion correction and reconstruction strategies are plotted. A lung nodule is pointed out (red dashed circle) in patient 1(5 y/o). A region of ground-glass opacity is shown (green dashed circle) in patient 2(4 y/o). Small pneumatoceles (lung cysts) are pointed out (blue arrow) in patient 3(8 y/o). Abnormality regions of all examples are zoomed in, shown in the second rows.

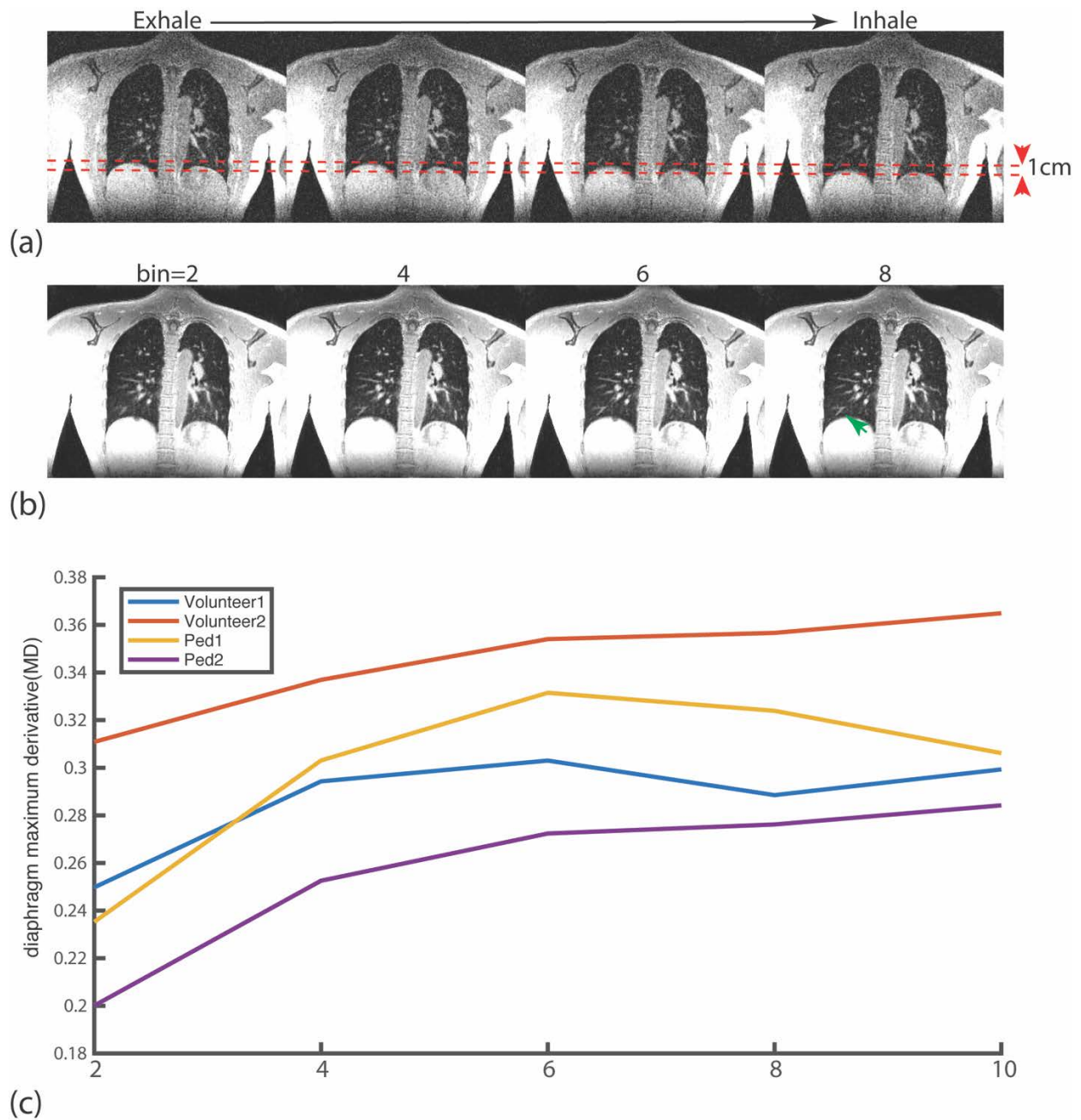

Figure 5. Effect of the number of motion states on the iMoCo reconstruction. An example in a healthy volunteer is shown. A motion resolved reconstruction (8 motion states) is shown in (a), where the diaphragm motion from exhale to inhale state is ~1cm. The iMoCo reconstruction with different motion states bin numbers are plotted in (b). Diaphragm maximum derivative(MD) are from 4 cases(2 from volunteers and 2 from pediatric patients) reconstructed via iMoCo using different number of motion states are plotted in (c).

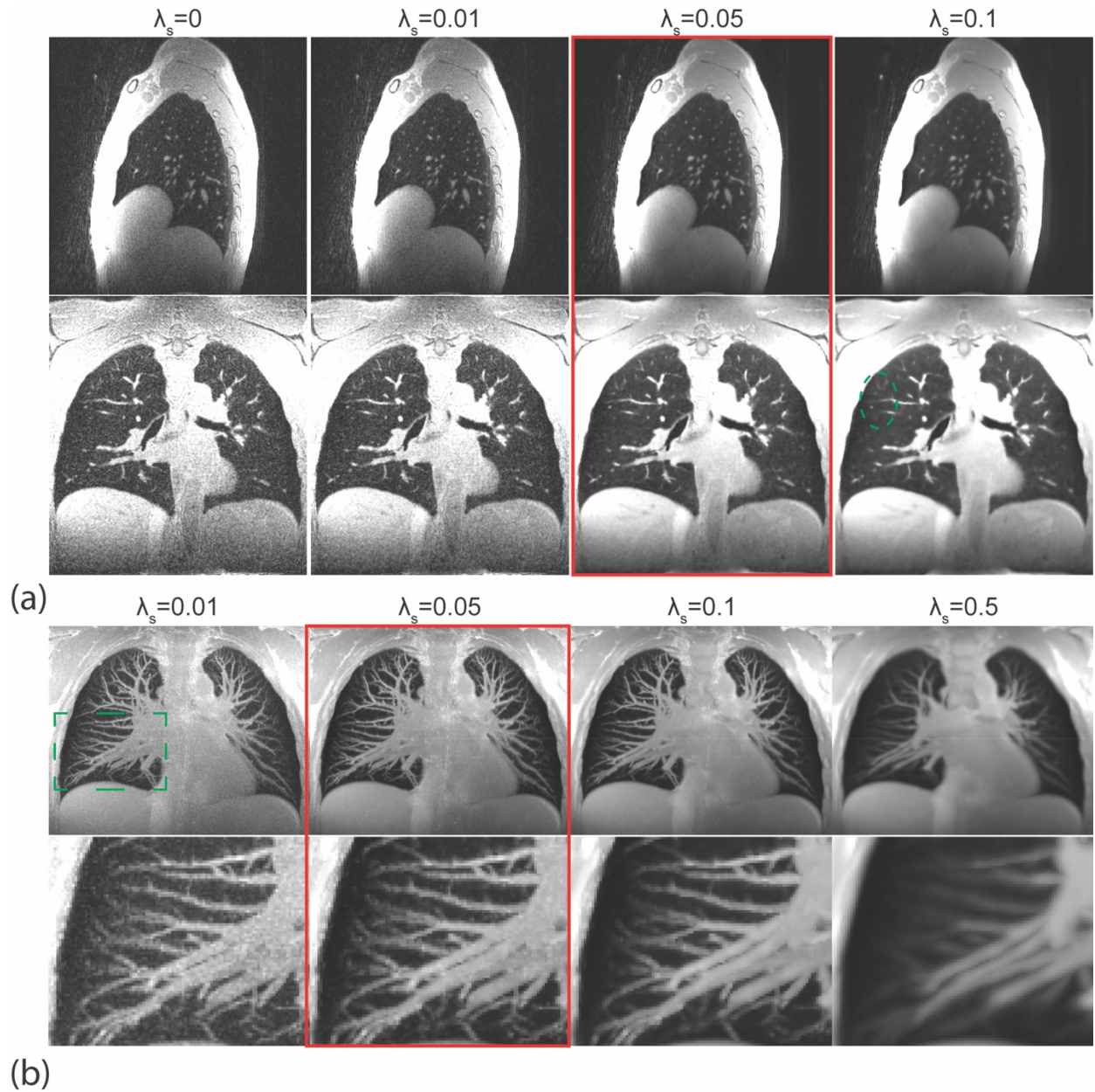

Figure 6. Effect of TGV sparse constraint parameter λ on the iMoCo reconstruction. A sagittal and coronal slice from a healthy volunteer with different TGV constraint parameter λ, from 0 to 0.1, are shown in (a). Maximum Intensity Projection(MIP) of 20 coronal slices of the reconstructed volume with different regularization levels are shown in (b), first row shows the MIP images, second shows the zoomed-in image of rectangular area in the first image.

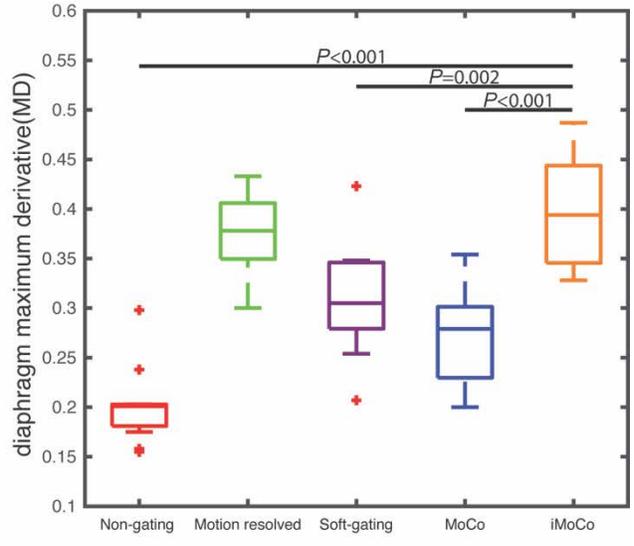

Figure 7. Diaphragm maximum derivative(MD) comparison, where a higher MD corresponds to a sharper edge. The MD calculation process is shown in (a). A MD comparison evaluated on 11 subjects (7 adult volunteers, and 4 pediatric patients) across different methods are plotted in (b).

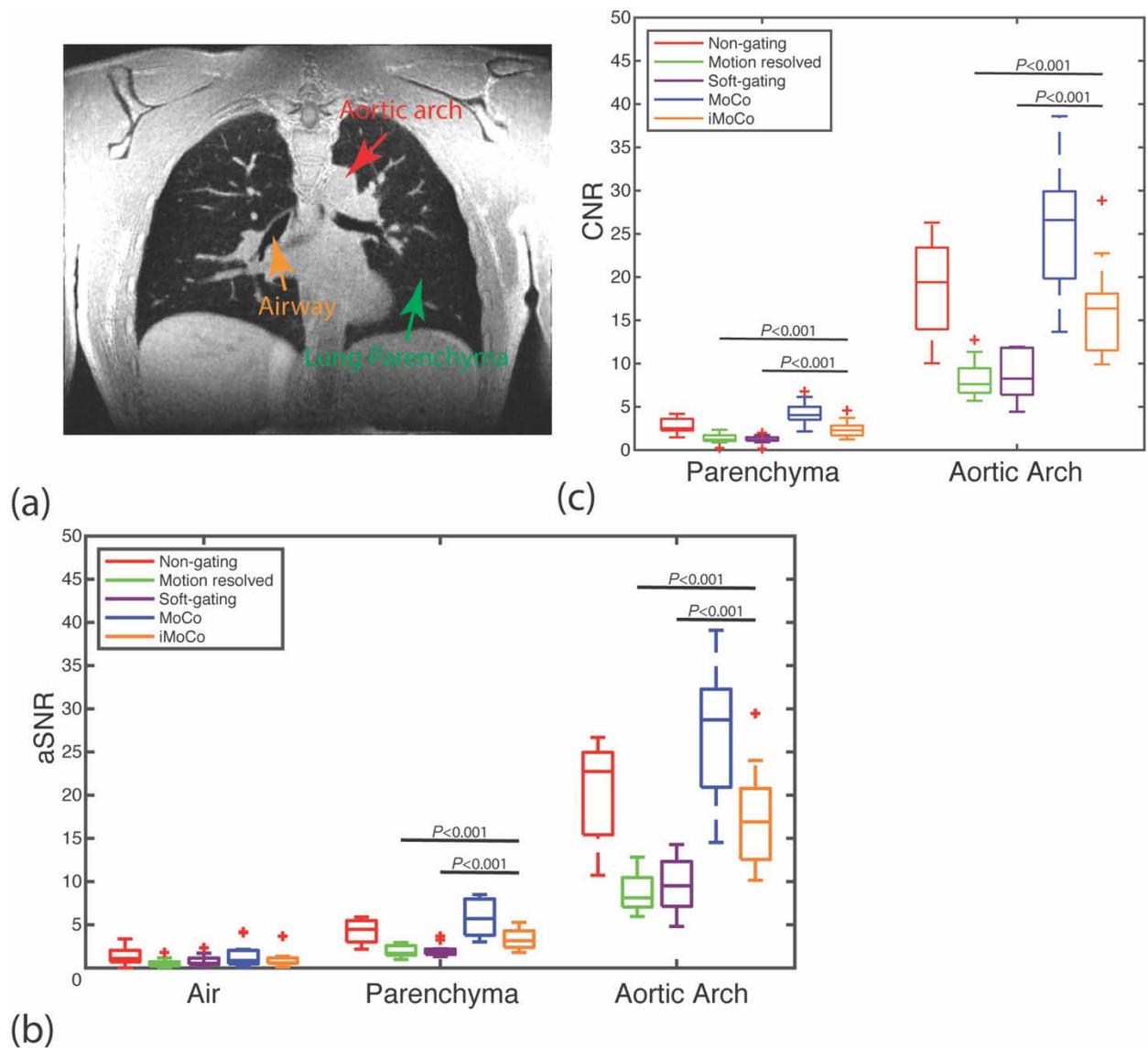

Figure 8. Apparent SNR(aSNR) and contrast-to-noise ratio(CNR) comparison between the reconstruction methods. Three different anatomical structures, a major airway, representative lung parenchyma ,and the aortic arch were manually annotated for aSNR and CNR measurements. An example of anatomical structures used for the measurements are shown in (a). Comparison of aSNR and CNR are separately plotted in (b), and (c). The increase in aSNR for the MoCo can be partially attributed to a loss of resolution due to smoothing in the reconstruction, which can be observed in Figs. 2-4. iMoCo had significantly higher parenchyma and aorta aSNR and CNR compared to motion-resolved and soft-gating reconstructions.

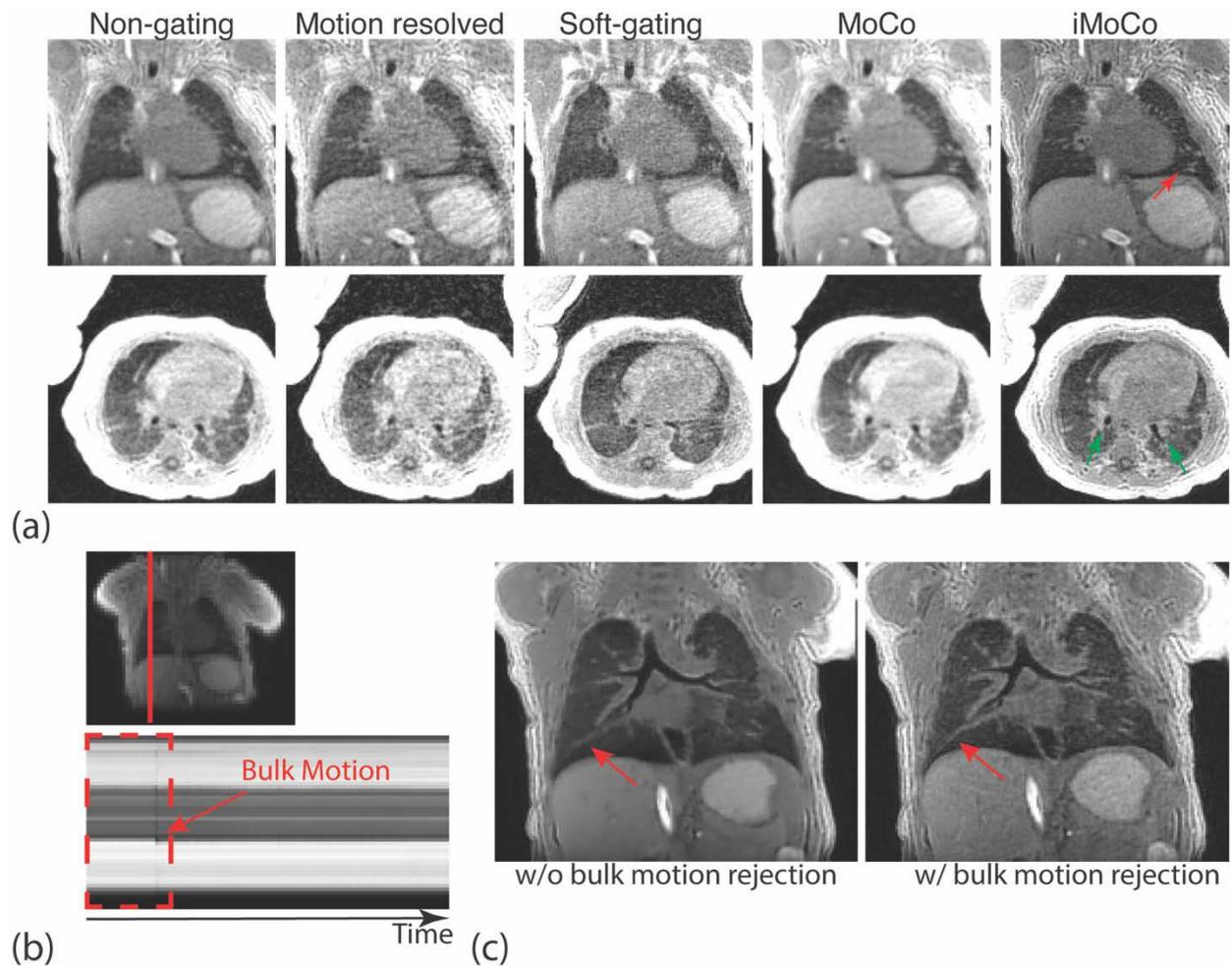

Figure 9. 10-week-old infant study results. (a) One coronal and one axial slice reconstructed with different motion correction and reconstruction strategies(with bulk motion rejection) are plotted. Vessels(red arrow) and airways(green arrows) are pointed out on the iMoCo images, showing improved delineation and contrast compared to other methods. (b) Bulk motion of the infant during the 5-minute scan was detected by an image based navigator, so data acquired prior to the bulk movement (red dashed rectangle time window) were rejected. (c) Images reconstructed by using iMoCo with and without bulk motion rejection are shown, where bulk motion rejection further reduces the motion effects.

**Supporting information:**

**Effect of motion resolved reconstruction regularization terms on motion field estimation**

We designed experiments to investigate how the regularization terms in the motion resolved reconstruction affect the motion estimation that is used for the motion compensation. Three different regularization combinations were used in this experiment, where the weightings for regularizations were empirically selected based on our previous experiments. All three reconstructions have motion dimension total variation regularizations (temporal TV) following the standard XD-GRASP[1] reconstruction, Eq (S1). Spatial total variation(TV) was added in one combination, Eq. (S2), while in another one spatial l1-wavelet regularization was added, Eq. (S3).

$$\underset{X}{\mathrm{argmin}} \sum_{i,k}^{N,m} ||W(FS_i X_k - d_{ik})||_2^2 + \lambda_t TV_t(X) \; . \tag{S1}$$

$$\underset{X}{\mathrm{argmin}} \sum_{i,k}^{N,m} ||W(FS_i X_k - d_{ik})||_2^2 + \lambda_s TV_s(X) + \lambda_t TV_t(X) \; . \tag{S2}$$

$$\underset{X}{\mathrm{argmin}} \sum_{i,k}^{N,m} ||W(FS_i X_k - d_{ik})||_2^2 + \lambda_s ||\Phi X||_1 + \lambda_t TV_t(X) \; . \tag{S3}$$

Notations used here are all the same as Eq. (1) in the main text. Note that Eq. (S3) is equivalent to Eq.(1) in the main text, and this formulation was used in the proposed iMoCo UTE method and all results in the main text.

After all reconstructions, the same non-rigid registration was used to estimation the motion field, and three different metrics were used to compare the result. Root Mean Square Error (RMSE) between two images:

$$\mathrm{RMSE} = \sqrt{\frac{\sum_{i=1}^{N}(I_1(i) - I_2(i))^2}{N}} \; . \tag{S4}$$

where $I_1(i)$ and $I_2(i)$ are $i$th voxels from $I_1$ and $I_2$. Here, we selected the expiratory state image as reference, the inspiratory state image as moving image. RMSE between two different states images was computed before and after image registration. RMSE was used to evaluate the overall registration performance.

Linear correlation coefficient(CC) and mean Euclidean distance($\overline{dist_E}$) between motion fields derived from different motion resolved reconstruction were computed to show the effects of regularization terms on the motion estimation step:

$$CC = \sum_{i=1,k=1}^{N,3} \frac{(u_1(i,k) - \overline{u_1})(u_2(i,k) - \overline{u_2})}{\sqrt{\sum_{i=1,k=1}^{N,3}(u_1(i,k) - \overline{u_1})^2 \sum_{i=1,k=1}^{N,3}(u_2(i,k) - \overline{u_2})^2}} \ . \tag{S5}$$

$$\overline{dist_E} = \frac{1}{N} \sum_{i=1}^{N} \sqrt{\frac{\sum_{k=1}^{3}(u_1(i,k) - u_2(i,k))^2}{3}} \ . \tag{S6}$$

Where $u_1(i,k)$ and $u_2(i,k)$ are two motion fields derived from different motion resolved reconstructions, $i$ is voxel index, and, since motion fields are 3D vectors in this study, therefore, $k$ is the motion vector index. The correlation coefficient shows the spatial similarity of two motion fields, and mean Euclidean distances measures the mean distance between the estimated motion fields. As there is no ground truth for motion estimation, we chose the group with only temporal TV regularization as reference for cross-correlation and mean Euclidean distance calculation.

According to Sup. Table S1, the RMSE of motion registration with different regularization is very similar. The correlation coefficient between motion fields estimated with different motion resolved reconstructions is close to 1, and mean Euclidean distances are all much smaller than 1 voxel, both of which indicate the motion field estimation is insensitive to the change of motion resolved reconstruction. Thus we do not expect these different combinations of regularization would affect the subsequent iterative motion compensated reconstruction.

**Effect of coarser resolution registration on iMoCo reconstruction**

**List of Figures and Tables**

Supporting Information Figure S1. Example of motion resolved reconstructions. In the first row are one end expiratory state image from the motion resolved reconstructions. In the second row are the estimated motion field in the S/I direction between the expiratory to inspiratory state.

Supporting Information Figure S2. Performance comparison among different registration methods. Three different registration methods, Demons, SyN with MSE similarity, and bspline, results are plotted in first row, (a). Difference between aligned image and reference image are plotted in second row, (a). Maximum Intensity Projection of the difference map between aligned image and reference image are shown in (b).

Supporting Information Figure S3. aSNRs in three different areas (lung parenchyma, aortic arch, and main airway) with different TGV regularization $\lambda_s$ levels from one of the cases shown in the Main Text, Figure 6.

Supporting Information Table S1. Comparison of registration RMSE with different motion resolved reconstruction regularization. As the registration RMSE depends on individual subject scan parameters, motion, etc., measurement of each subject is separately listed. RMSE between two states images before registration is calculated for reference.

Supporting Information Table S2. Summary of correlation coefficient and mean Euclidean distance of motion fields with different regularizations. As there is no ground truth for motion estimation, we chose the group with only temporal TV regularization as reference for correlation coefficient and mean Euclidean distance calculation. The two metrics are measured in all (N = 11) subjects, and the mean and standard deviation are listed.

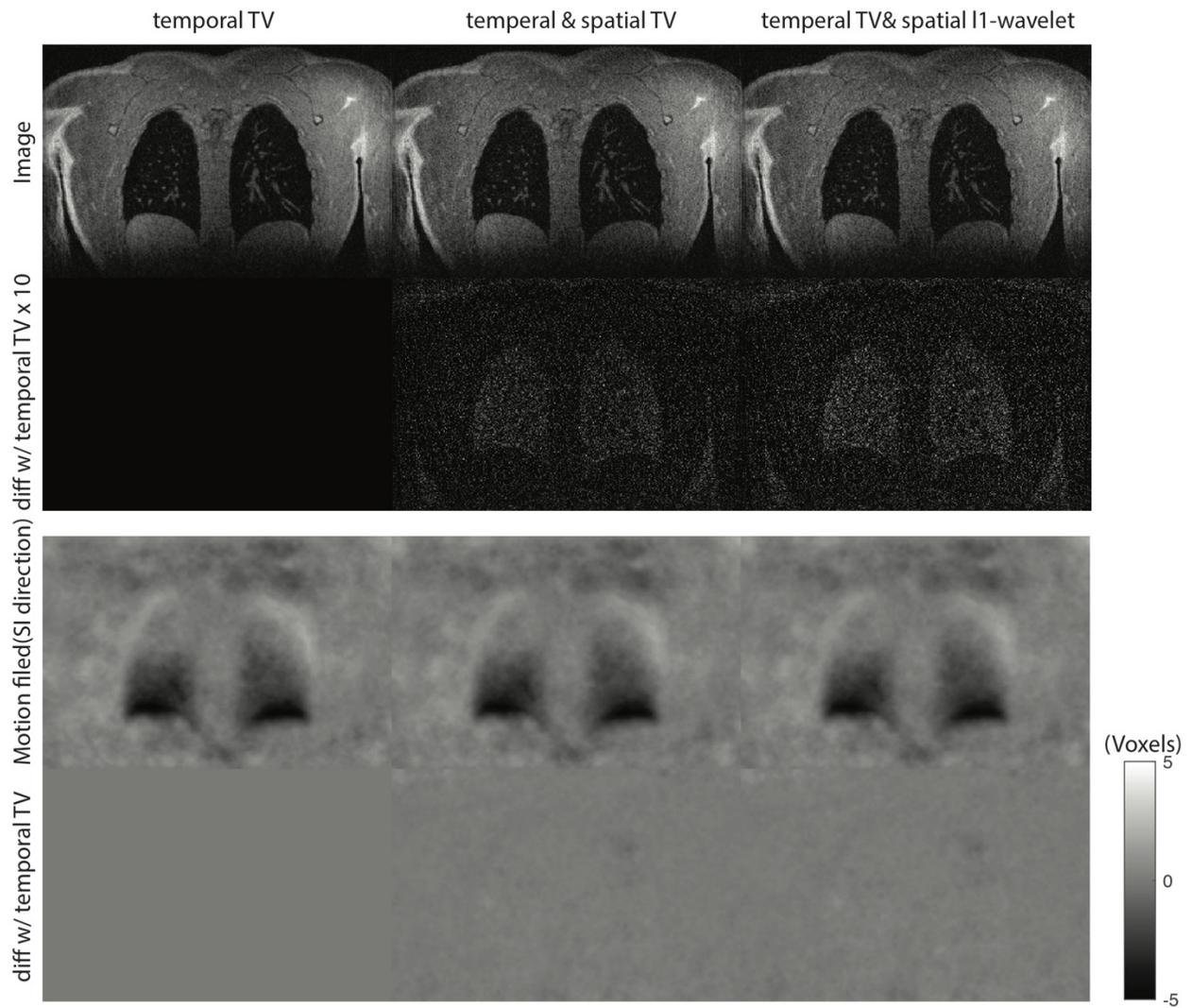

Supporting Information Figure S1. Example of motion resolved reconstructions. In the first row are one end expiratory state image from the motion resolved reconstructions. In the second row are the estimated motion field in the S/I direction between the expiratory to inspiratory state.

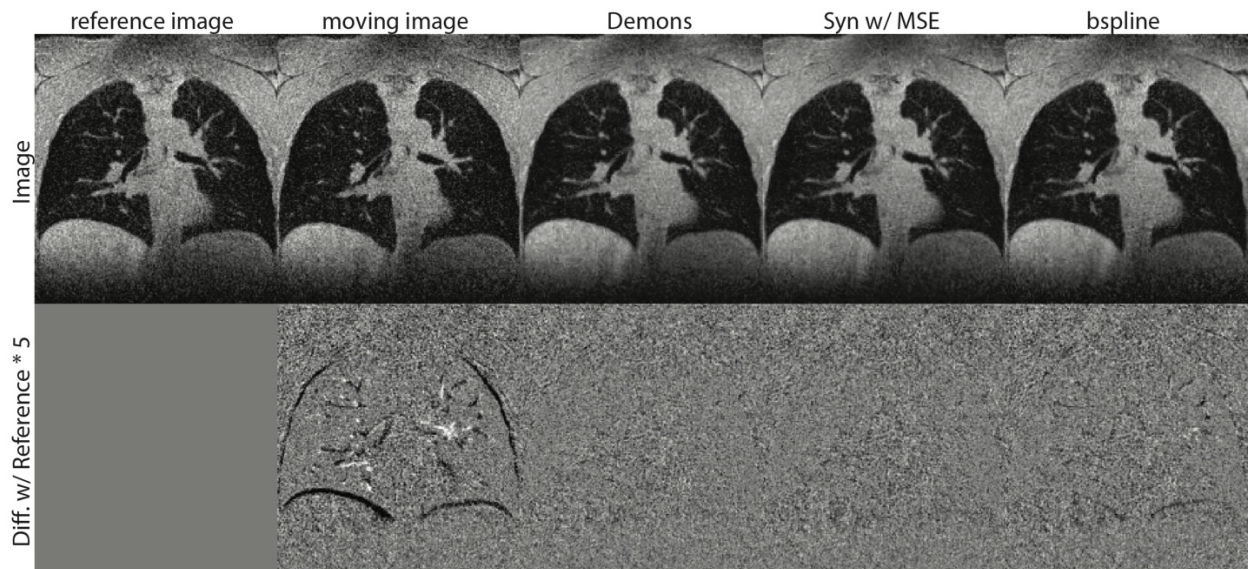

(a)

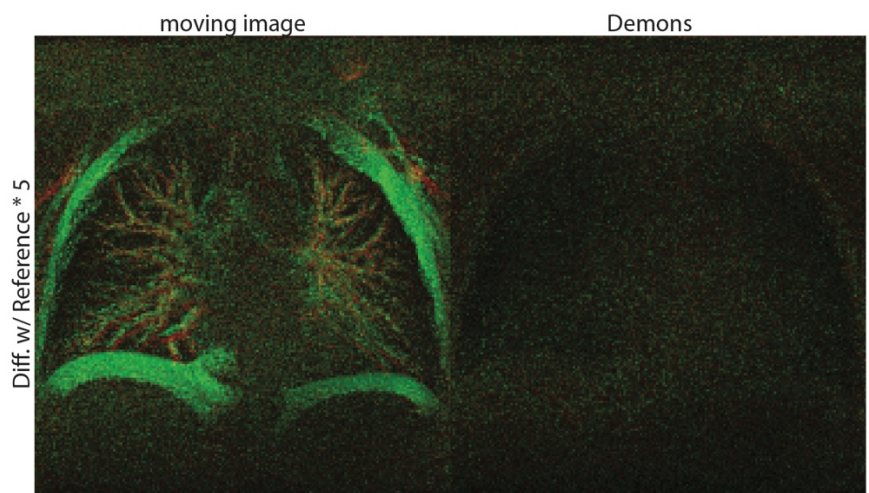

(b)

Supporting Information Figure S2. Performance comparison among different registration methods. Three different registration methods, Demons, SyN with MSE similarity, and bspline, results are plotted in first row, (a). Difference between aligned image and reference image are plotted in second row, (a). Maximum Intensity Projection of the difference map between aligned image and reference image are shown in (b).

| Subject No. | w/o registration | Temporal TV | Temporal & spatial TV | Temporal TV & spatial l1-wavelet |
|---|---|---|---|---|
| 1 | 29.89 | 20.12 | 19.83 | 20.05 |
| 2 | 50.76 | 47.89 | 47.64 | 47.75 |
| 3 | 58.75 | 32.86 | 32.55 | 32.78 |
| 4 | 51.74 | 34.54 | 34.16 | 34.43 |
| 5 | 57.92 | 29.58 | 29.20 | 29.52 |
| 6 | 62.82 | 39.09 | 38.51 | 38.98 |
| 7 | 38.16 | 25.93 | 25.56 | 25.85 |
| 8 | 79.42 | 53.84 | 52.98 | 53.61 |
| 9 | 69.44 | 35.12 | 34.83 | 35.03 |
| 10 | 53.61 | 30.12 | 29.82 | 30.08 |
| 11 | 103.43 | 49.76 | 49.89 | 49.87 |

Supporting Information Table S1. Comparison of registration RMSE with different motion resolved reconstruction regularization. As the registration RMSE depends on individual subject scan parameters, motion, etc., measurement of each subject is separately listed. RMSE without registration is calculated for reference.

| N = 11 | Temporal & spatial TV | Temporal TV & spatial l1-wavelet |
| --- | --- | --- |
| **correlation coefficient** | 0.95±0.02 | 0.99±0.01 |
| **mean Euclidean distances(voxels)** | 0.20±0.04 | 0.07±0.04 |

Supporting Information Table S2. Summary of correlation coefficient and mean Euclidean distance of motion fields with different regularizations. As there is no ground truth for motion estimation, we chose the group with only temporal TV regularization as reference for correlation coefficient and mean Euclidean distance calculation. The two metrics are measured in all (N = 11) subjects, and the mean and standard deviation are listed.

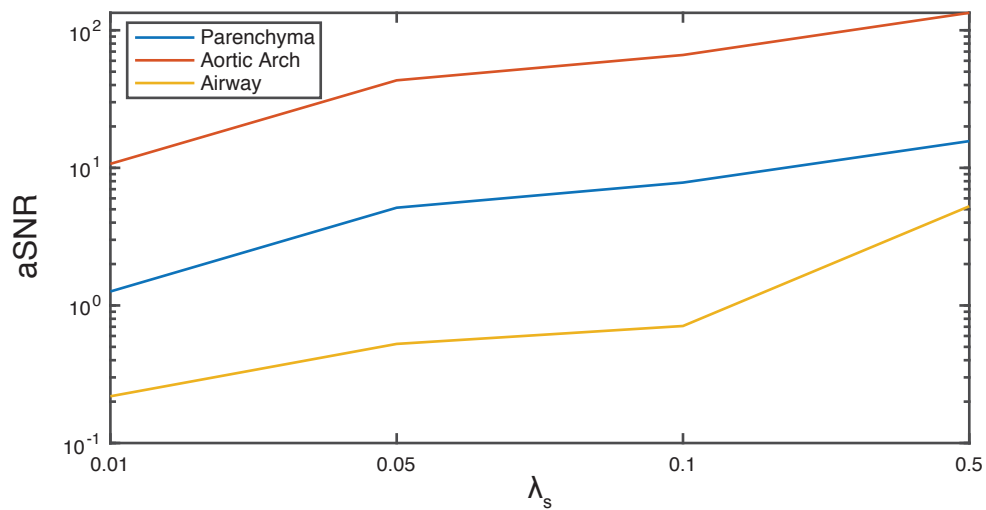

Supporting Information Figure S3. aSNRs in three different areas (lung parenchyma, aortic arch, and main airway) with different TGV regularization $\lambda_s$ levels from one of the cases shown in the Main Text, Figure 6.